\documentstyle[preprint,aps]{revtex}
\tightenlines
\begin{document}

\title{Properties of resonant activation phenomena}

\author{Mari\'{a}n Bogu\~n\'{a}, Josep M. Porr\`{a}, Jaume Masoliver} 
\address{
Departament de F\'{\i}sica Fonamental,
Universitat de Barcelona, Diagonal 647, 08028-Barcelona}
\author
{Katja Lindenberg}
\address{Department of Chemistry and Biochemistry 0340
and Institute for Nonlinear Science,
University of California San Diego,
La Jolla, California 92093-0340}

\date{\today}
\maketitle

\begin{abstract}
The phenomenon of resonant activation of a Brownian particle
over a fluctuating barrier is revisited.  We discuss the important
distinctions between barriers that can fluctuate among ``up" and ``down"
configurations, and barriers that are always ``up" but that can fluctuate
among different heights.  A resonance as a function of the barrier
fluctuation rate is found in both cases, but the nature and physical
description of these resonances is quite distinct. 
The nature of the resonances, the physical basis for the resonant
behavior, and the importance of boundary conditions are discussed
in some detail.  We obtain analytic
expressions for the escape time over the barrier that explicitly capture
the minima as a function of the barrier fluctuation rate, and show that
our analytic results are in excellent agreement with numerical results.
\end{abstract}


\section{Introduction}

Noise-induced non-equilibrium phenomena in nonlinear systems have recently
attracted a great deal of attention in a variety of contexts \cite{Millonas}.
In general, these phenomena involve a response of the system that is not only
produced or enhanced by the presence of the noise, but that is {\em
optimized} for certain values of the parameters of the noise. One example is
the phenomenon of stochastic resonance \cite{Bulsara}, wherein the response
of a nonlinear system to a signal is enhanced by the presence of noise and
maximized for certain values of the noise parameters.  Another involves
nonequilibrium ratchets, wherein intrinsically unbiased Brownian motion in
stochastic asymmetric potentials leads to a systematic drift motion whose
magnitude and even direction can be tuned by the parameters of the noise
\cite{Magnasco,Wenow}. A third is the recent discovery of a re-entrant
noise-induced phase transition in a nonlinear coupled array, that is, a
transition that is only observed for certain finite ranges of noise
parameters \cite{Broeck}. A fourth such phenomenon, the one of interest to us
in this paper, has been called ``resonant activation'' and was first
identified by Doering and Gadoua \cite{DG} and further studied by a
number of other authors.  Here the mean escape time of a particle driven
by (usually white) noise over a
barrier of randomly fluctuating height exhibits a minimum as a function of
the parameters of the barrier fluctuations.

Our interest in this problem first arose because it seemed to us
that for sufficiently simple potentials it should be possible to find
{\em analytic} dependences of the escape rate on the system parameters
(or at least good approximations to them) and, more specifically,
that it should be possible to
find analytic expressions for the parameter combinations that lead
to the minimum in the escape rate.  Some analytic results are
available \cite{DG,bier,DN,OU,iwan,marchi},
including those in the original work of Doering and Gadoua that apply to
a very specific circumstance discussed in more detail below. In general,
however, most available results are numerical\cite{marchi}.
Analytic results are scarce, and usually apply only to one parameter
regime {\em or} another and are thus unable to explicitly
capture the occurrence of a minimum in the escape rate.  A minimum
in the escape rate usually arises from
these approximations only by inference, and the approximations provide
no way to specifically locate the minimum except as an intersection
point of two unrelated approximations.  They also do not provide a way
to determine the dependence of the minimum on the system parameters.

In this paper we accomplish our goal, that is, we obtain a number of
analytic results for moments of the first passage time over a fluctuating
barrier for the particular model system (a triangular potential
barrier subject to dichotomous fluctuations)
used in a number of studies of resonant activation. In particular,
we obtain analytic approximations that explicitly capture not only
the minimum in the escape rate but that allow us to study the
variability of the escape rate in parameter space, that is, the
depth and width of this minimum.

In the process of obtaining these results we have also accomplished a
number of important clarifications on the nature of models that
have been presented under the common ``resonant
activation'' rubric, and on the nature of resonant activation itself. 
Some of these models in fact differ from
one another in essential respects.  We discuss these clarifications
and differences in some detail, and thus shed some light on the role
played by the interplay of the white noise and the barrier fluctuations
on the escape process. 
We anticipate some of our findings:
\begin{itemize}
\item
A distinction must be made between situations in which the
fluctuating barrier can be ``up" or ``down" (i.e. can go from being a
barrier to being flat or even a well), and situations in which there is
always a barrier.  Although resonant behavior can be observed in all
cases, the physical picture underlying this behavior is different in
different cases.
\item
Boundary conditions play an extremely important role in the problem.
\item
The qualitative physical description of the resonance in the fluctuating
barrier problem is as follows.  When the barrier fluctuates extremely
slowly within a range $0< \upsilon_{min}\leq \upsilon
\leq  \upsilon_{max}$, the mean first passage
time to the top of the barrier is extremely long because it is dominated
by those realizations for which the barrier starts in the high position.
The mean first passage time is then proportional to $e^{\upsilon_{max}/D}$
where $D$ is the intensity of the white noise.
Indeed, if the barrier fluctuation rate is smaller than the inverse of the
mean first passage time to the highest barrier, the barrier is essentially
quasi-static throughout the process.  At the other extreme, if the barrier
fluctuates very rapidly, the mean first passage time is determined by the
average barrier $\upsilon_0$, i.e. it is proportional to
$e^{\upsilon_0/D}$.  Between these extremes, and over a broad range of
barrier fluctuation rates, passage over the barrier occurs primarily when
the barrier is low, and the mean first passage time is then proportional
to $e^{\upsilon_{min}/D}$.  This dependence is quite robust, and the
prefactor determines the actual minimum within this broad range.
\item
This behavior does not require that the barrier
fluctuate; an oscillatory variation of the barrier height yields
essentially the same results.
\end{itemize}

In Sec.~\ref{statement} we provide a detailed statement of the resonance
activation problem.
Section~\ref{doergad} discusses the analytic solution of the ``up"--``down" 
case; we show that the resonance flipping rate and the resonance
activation in this case are independent of the white noise intensity.
In Sec.~\ref{simpler} the significance of the white noise
and of the boundary conditions in this ``up"--``down"
problem are discussed in detail. Section~\ref{bierastum} deals with the
case of barrier fluctuations when the barriers are always high. We 
obtain a single analytic expression for the mean first passage time that
exhibits a minimum as a function of the barrier fluctuation rate and
that in fact quantitatively captures the correct behavior over most of
parameter space, as determined by comparison with numerical results.
With this result we are able to determine the resonance frequency
analytically, and also the range of barrier fluctuation rates over which the 
mean first passage time is essentially flat.  In Sec.~\ref{square} we
discuss the case of a barrier that oscillates (rather than fluctuates).
This case also exhibits resonant activation, although some of the
quantitative details of the problem are slightly modified.  Finally, we
conclude with a summary and some final points in Sec.~\ref{conclusions}.

\section{Statement of the Problem}
\label{statement}
Consider a process that evolves in a bistable
potential and is driven by weak
Gaussian white noise, so that the process is
occasionally able to cross
from one minimum of the bistable potential to the other.  If the
parameters of the system are fixed in time, the rate at which the
process crosses from one well to the other under a variety of conditions
is well known (e.g. the Kramers rate).  Suppose now that the height of
the barrier separating the two minima of the bistable potential
fluctuates in time.  
We wish to explore the effect of the barrier fluctuations on the rate
of passage of the process from one well to the other.  
More specifically, it is known that there is an
optimal barrier fluctuation rate that minimizes the passage time from
one well to the other for given parameter
values \cite{DG,bier,DN,OU,iwan,marchi},
This minimum identifies the phenomenon of {\em resonant activation}. 
We are interested in the analytic properties of the resonant
activation phenomenon.
Note that the barrier fluctuations here are such that the energy
difference between the potential minima remains constant --- only the
barrier {\em height} fluctuates.  This is to be contrasted with the
phenomenon of stochastic resonance, where the energy difference is
modulated by a small periodic signal.   

We adhere to the overdamped regime, and hence the process
$y(\tau)$ evolves according to the Langevin equation
\begin{equation}
	\dot{y}(\tau)=-V'(y)-g'(y)\eta(\tau)+\xi(\tau).
  \label{basiceq}
\end{equation}
Here $\xi(\tau)$ is zero-centered Gaussian white noise
with correlation function
\begin{equation}
		\langle \xi(\tau) \xi(\tau ') \rangle =2D \delta(\tau-\tau ').
	\label{white}
\end{equation}
One can think of the white noise as arising from a heat bath, in which case
the diffusion coefficient $D$ is proportional to the bath temperature $T$.
Time is measured in units of the friction coefficient,
which has been set to unity in Eq.~(\ref{basiceq}).

The potential $V(y)$ is a bistable potential,
typically with isoenergetic minima.  Doering and Gadoua \cite{DG}
introduced the
triangular potential shown in Fig.~\ref{fig1}.  The potential barrier is
defined by
\begin{equation}
V(y)=\left\{
\begin{array}{cc}
\upsilon_0y/L &  \quad 0 \leq y <L \\
-\upsilon_0y/L+2\upsilon_0 &\quad  L\leq y \leq 2L 
\end{array}
\right. 
\label{potential}
\end{equation}
and the potential rises to infinity at $y=0$ and at $y=2L$. In the
absence of the contribution $g'(y)\eta(\tau)$ in Eq.~(\ref{basiceq}) this
represents a standard problem where the rate $\kappa$ at which the process
crosses the barrier at $x=L$ is related to the mean first passage
time $\widetilde{T}_1$ from the bottom of one of the
wells, say the one at $y=0$, to the top of the barrier:
$\kappa=1/2\widetilde{T}_1$. 
To calculate the mean first passage time one assumes a reflecting
boundary condition at
$y=0$ and an absorbing boundary condition at $y=L$.

In the resonant activation problem we have, in addition, the
contribution $g'(y)\eta(\tau)$. Here $\eta(\tau)$ is a
non-equilibrium noise
that coupled to $g'(y)$ causes the potential barrier to fluctuate. It is
a nonequilibrium noise because there is no dissipative
contribution in the equation of motion associated with this fluctuating
term, and hence the system is open.
The noise $\eta(\tau)$ is usually taken to be
exponentially correlated, the most ubiquitous choices being 
Ornstein-Uhlenbeck noise \cite{OU,iwan,marchi} and 
Markovian dichotomous noise \cite{DN,marchi}.  Here we deal only with
the latter: $\eta(\tau)$ takes on the values
$\pm 1$, and  the change from one to the other is distributed in time
according to the exponential density function
\begin{equation}
\phi(\tau)=\gamma e^{-\gamma \tau}
\label{flips}
\end{equation}
so that the flipping rate of the dichotomous noise is $\gamma$. 
The fluctuating barrier is accomplished by picking for $g(y)$ the function
\begin{equation}
g(y)=\left\{
\begin{array}{cc}
\alpha y/L &  \quad 0 \leq y <L \\
-\alpha y/L+2\alpha & \quad L\leq y \leq 2L 
\end{array}
\right. 
\label{g}
\end{equation}
and zero otherwise.
The addition of the random potential term $g(y)\eta(\tau)$
causes the potential
barrier to switch between the two values $\upsilon_+=
\upsilon_0+\alpha$ and $\upsilon_-=\upsilon_0-\alpha$.

We wish to calculate the rate at which the process crosses the
point $y=L$, which in turn is related, as before, to
the mean first passage time $\widetilde{T}_1$ from $y=0$ to $y=L$
when a reflecting
boundary is located at $y=0$ and an absorbing boundary at $y=L$. 
(The distortions in the potential profile that may be caused by multiplicative
noise, and the implications on the appropriate definition of an escape
time, are well known and have been widely discussed in the literature;
see. e.g. \cite{marchi}.  The potentials used here do not exhibit such
distortions.)
In particular, we wish to establish analytically the dependence
of $\widetilde{T}_1$ 
on the flipping rate $\gamma$ and to identify the flipping rate for
which $\widetilde{T}_1$ is a minimum.

Doering and Gadoua \cite{DG} calculated the mean first passage time for
this model
in the absence of the potential $V(y)$, that is, when the ``barrier'' 
flips between being a true barrier
(of height $+\alpha$) to being a well ($-\alpha$), and they
obtained a resonance phenomenon, that is, the mean first passage
time from $0$ to $L$ exhibits a minimum at a particular value of the
flipping rate $\gamma$.  Doering and Gadoua also presented simulation
results
for the case $\upsilon_0=\alpha$, that is, when the ``barrier'' flips
between being
a true barrier (height $2\alpha$) and there being no barrier.  Bier and
Astumian \cite{bier} considered the true barrier case, that is, the case
where there is always a barrier (in fact,
they took $\upsilon_0\gg \alpha$) and obtained numerical results that show a
resonance.  Their analytic barrier crossing rate results are
obtained separately for low
flipping rates (small $\gamma$) and for high flipping rates
(large $\gamma$). Neither
result in itself exhibits a minimum, although one can infer the
existence of a minimum (but not its dependence on the system parameters;
see also \cite{marchi}) from their combination.

With this general statement of the problem we can be more precise about
the results that we present in this paper.  First, we consider the
$\pm \alpha$ barrier-well case of Doering and Gadoua, reproduce their
analytic results for the mean first passage time,
and also obtain analytic results for the resonant mean first passage
time, the resonant flipping rate, and for the second moment of the first
passage time distribution.  We argue that the $\pm \alpha$ barrier
case represents a situation that is
completely different from the ``true barrier'' case
considered by Bier and Astumian. 
Both exhibit resonance behavior, but via different mechanisms. 
We explore these differences and interpret the Doering--Gadoua case on the
basis of an even simpler model.  Furthermore, we obtain analytic
results for the high barrier case considered by Bier and Astumian
that yield an {\em explicit} minimum in the mean first passage time as a
function of the flipping rate.  We present analytic results for the
mean first passage time at resonance, for the resonant flipping rate,
and we analyze the behavior of the system away from this point to
assess how sharp this resonance might be.

We shall present our analysis and results in terms of the dimensionless
variables
\begin{equation}
t\equiv \tau D/L^2, \qquad x\equiv y/L, \qquad T_1 \equiv
\widetilde{T}_1 D/L^2,
\label{variables}
\end{equation}
and the dimensionless parameters
\begin{equation}
a\equiv \alpha/D,\qquad V_0\equiv \upsilon_0/D,
\qquad \lambda\equiv \gamma L^2/D,
\label{parameters}
\end{equation}

The differential equation for the mean first passage time in all
cases considered in this paper is given by
\begin{equation}
\frac{d^4T_1}{dx^4} -2V_0 \frac{d^3T_1}{dx^3}
+(V_0^2 - a^2-2\lambda)\frac{d^2T_1}{dx^2} +2\lambda V_0
\frac{dT_1}{dx} = 2\lambda
\label{t1}
\end{equation}
with the boundary conditions
\begin{equation}
\left.\frac{dT_1}{dx}\right|_{x=0}   = 0,
\label{con1}
\end{equation}
\begin{equation}
\left. \frac{d^2T_1}{dx^2}\right|_{x=0} = -1,
\label{con2}
\end{equation}
at the reflecting boundary and
\begin{equation}
T_1(x=1) = 0,
\label{con3}
\end{equation}
\begin{equation}
\left[\frac{d^3T_1}{dx^3}
- 2V_0 \frac{d^2T_1}{dx^2} +(V_0^2-a^2)
\frac{dT_1}{dx}\right]_{x=1} = V_0
\label{con4}
\end{equation}
at the absorbing boundary.

A brief description of how this equation and boundary conditions
arise is given in the Appendix.

\section{Analytical solution of the Doering--Gadoua Model}
\label{doergad}

Consider the mean first passage time to $x=1$ ($y=L$)
when the mean barrier height is $V_0=0$. The solution as a
function of the initial position
$x$ can in this case be given analytically:
\begin{eqnarray}
	T_1(x)
	&=& (x-1)\left[ \frac{2 \lambda a^2}{\mu^3}
		\frac{\mu - \sinh(\mu)}{a^2+2 \lambda
		\cosh(\mu)}-\frac{\lambda}{\mu^2 }(x+1) \right]
		- \frac{2 a^2}{\mu^4}\frac{ \sinh[\mu (x-1)/2]}
		{a^2+2 \lambda \cosh(\mu )}
		\nonumber\\
		&&\nonumber\\
	& \times & \left\{ a^2 \sinh[\mu (x+1)/2] +
	2 \lambda \sinh[ \mu (x-1)/2] 
		+2 \mu \lambda \cosh[ \mu (x+1)/2] \right\},
\label{t1sim}
\end{eqnarray}
where we have introduced the symbol
\begin{equation}
\mu=\sqrt{a^2+2\lambda}. 
\end{equation}
This result has been previously reported for the particular initial value
$x=0$\cite{DG}.  We know from Ref.~\cite{DG} that $T_1(x=0)\equiv
T_1$ exhibits a
resonance with respect to $\lambda$; we wish to establish the resonance
flipping rate $\lambda_{res}$ and the behavior of the mean first passage time
at this resonance point.

The expression for the mean first passage time simplifies considerably when
the dimensionless quantity $\mu$ is large ($\mu \gg 1$), which is
the physically interesting weak-white-noise regime. Indeed, the only way
that $\mu $ could be small is if $\alpha$ and $\lambda$ are small, that is,
if (in dimensioned units) the white noise intensity $D$ is 
greater than the barrier height $\alpha$ and greater
than $2\gamma L^2$.  The first condition renders the problem uninteresting
- if barriers are on average lower than the noise then one has an essentially
free diffusion problem.  The second condition requires a small system with
a low flipping rate, again a very specific situation that is not
particularly interesting in this context.  The customarily interesting
physical situation occurs when the white noise is weak compared to
the barrier
height, that is, when $a \gg 1$ and this in turn leads to
$\mu \gg 1$.
We use these two statements of the ``interesting regime'' interchangeably.

When $\mu$ is sufficiently large the following approximations
are valid:
\begin{equation}
	\cosh \mu  -1 \sim \cosh \mu \sim  \sinh \mu  - \mu  \sim 
         \sinh \mu  \sim \frac{1}{2} e^{\mu}. 
\end{equation}
If in addition
\begin{equation}
\lambda \gg a^2e^{-a}
\label{ll}
\end{equation}
then the result simplifies even further, and one finally obtains the
following much simpler approximate expression:
\begin{equation}
	T_1 (x) \sim \frac{a^2(a^2- 2\lambda e^{-\mu x})}
	{2 \lambda (a^2+2\lambda )^2}
		+ \frac{a^2(2-x)}{(a^2+ 2\lambda )^{3/2}}
		+ \frac{ \lambda (1-x^2)}{a^2+ 2\lambda}
		+O(a^2e^{-\mu}).
	\label{2.17}
\end{equation}
It can be shown that this expression as a function of $\lambda$ has a minimum
at a finite value $\lambda_{res}$
that to leading order is of the form
$\lambda_{res} \sim a$.  This minimum is identified among the
roots of $dT_1(x)/dx=0$ as the one that coincides with
the minimum of the
complete expression of $T_1(x)$, Eq.~(\ref{t1sim}), as
$a\rightarrow\infty$.  Explicitly, following \cite{DG} we set
$x=0$ to further simplify the analysis: 
\begin{equation}
	T_1 \sim \frac{(a^2- 2\lambda)a^2}{2 \lambda 
		(a^2+2\lambda)^2}
		+ \frac{2a^2 }{(a^2+ 2\lambda)^{3/2}}
		+ \frac{ \lambda}{a^2+ 2\lambda}+O(a^2e^{-\mu }).
\label{r1}
\end{equation}
The extrema of $T_1$ as a function of $\lambda$ obey the equation 
\begin{equation}
	\frac{\lambda^2 }{a^2} (a^2+ 2\lambda) -
	\frac{2\lambda}{a^2} (a^2- 2\lambda ) 
	-\frac{1}{2} (a^2+2\lambda) -\frac{6\lambda^2}{a^2}
    \left(a^2 + 2\lambda\right)^{1/2} =0.
	\label{2.18}
\end{equation}
This equation can be solved perturbatively by taking
$ \lambda \sim a\left(\lambda_0
+\lambda_1 a^{-1} +\lambda_2 a^{-2} \right)$,
an expansion consistent with the fact that $a \gg 1$. Once this
expansion is substituted in
Eq.~(\ref{2.18}), the following result is obtained for the resonant
flipping rate as a function of $a$:
\begin{equation}
	\lambda_{res} \sim \frac{1}{\sqrt{2} }a 
		+(1+\frac{3}{\sqrt 2} )
	+ \frac{3}{2} (3+\frac{7}{\sqrt2 }) \frac{1}{a}, \qquad
	a\gg 1.
\label{rres}
\end{equation}
Note that the condition $\lambda\gg a^2e^{-a}$ is thus
satisfied.  The value of $T_1$ at the resonant flipping rate is
\begin{equation}
	T_1(\lambda=\lambda_{res})\equiv T_{res} \sim  
	\left( 2+  \sqrt{2} \right) \frac{1}{a}
	-\left(4+3\sqrt{2} \right) \frac{1}{a^2} 
	+\left(\frac 32+\frac 1{\sqrt{2}}\right) \frac{1}{a^3} 
\label{r2}
\end{equation}
provided $a$ is large.
For small values of $ \lambda $ ($\lambda \ll a^2 e^{-a}$) one
readily finds from Eq.~(\ref{t1sim}) with $x=0$ that
\begin{equation}
	T_1 \sim \frac{1}{2a^2} e^{a} - \frac{\lambda}{2a^4} e^{2a} 
	+\frac{\lambda^2}{2a^6}e^{3a}+ O(\frac{\lambda^3}{a^6}e^{4a}),
\label{r3}
\end{equation} 
when $a \gg 1$.

In Fig.~\ref{fig2}, the exact expression of $T_1$ is
compared with
the approximation (\ref{r3}) when $\lambda \rightarrow 0$ and with
the approximation (\ref{r1}) when $\lambda \rightarrow \infty$.

The second order moment of the first passage time distribution,
$T_2, $ also exhibits a resonance.
However, the resonant frequency of $T_2$ does not coincide with that
of $T_1$. This
means that there does not exist a unique resonant frequency or universal
scaling associated with the first passage time distribution.
$T_2$ can be calculated in a similar way (albeit even more expansively)
than $T_1$.  However, the full expression for $T_2(x)$ is too long
to be included here.
Instead, we only reproduce the expression for $T_2$ at $x=0$ when $a\gg 1$,
\begin{equation}
	T_2 \sim \frac{a^4(a^4+a^2\lambda-4 \lambda^2)}{\lambda^2 \mu^8}+
     \frac{4a^2(a^2-\lambda^2)}{\lambda \mu^7}+
     \frac{2a^2(4a^2-\lambda)}{\mu^6}+
     \frac{20\lambda a^2}{3 \mu^5}+
     \frac{5\lambda^2}{3\mu^4}+
     O(a^4e^{-\mu}).
\label{t2}
\end{equation}
In Fig.~\ref{fig3} this approximate result is
compared with the exact expression for $T_2$ when $a=8$. The behavior
of the resonant frequency $\lambda_{res,2}$ of $T_2$
can be obtained from the
expression for $T_2$ when $a\rightarrow \infty$. The calculation yields 
\begin{equation}
	\lambda_{res,2}\sim b_0 a +b_1+O(a^{-1}),
\end{equation}
where $b_0$ is the positive real solution of the equation
$5 b_0^4 + 10b_0^3 - 6 b_0+ -3=0$.
Numerical solution of this equation leads to the value
$b_0=0.825724\ldots$. With this value, the coefficient $b_1$ can also be
evaluated numerically and one obtains $b_1=3.56057$. Note that
$\lambda_{res,2} > \lambda_{res}$. Finally,
$T_2$ at the resonant frequency behaves as $1/a^2$ when $a\rightarrow
\infty$: 
\begin{equation}
  T_2(\lambda_{res,2})\sim \frac{c_0}{a^2}+\frac{c_1}{a^3}+O(a^{-4}),
\end{equation}
with $c_0=20.9521\ldots$ and $c_1=-104.244\ldots$.

Equations (\ref{r1}), (\ref{rres}), (\ref{r2}), and (\ref{r3}) are
the main results of
this section - they give the mean first passage time as a function of the
barrier height (or well depth) $a$, provided $a\ll e^a$, for all values of
$\lambda$.  The resonant behavior of the mean first passage time as a
function of the flipping rate of the barrier is clear and dramatic. 
It is useful to exhibit explicitly the limiting results in the original
units so that the dependence on system parameters is clear.  
When the flipping rate is large ($\lambda \gg a^2$ or
$\gamma \gg \alpha^2/DL^2$) the mean first
passage time from $y=0$ to $y=L$ grows as
\begin{equation}
\widetilde{T}_1 \rightarrow \widetilde{T}_{white} = \frac{L^2}{2D_{eff}},
\label{d1}
\end{equation}
as seen from Eq.~(\ref{r1}).  Here
\begin{equation}
D_{eff}\equiv D+\frac{\alpha^2}{2\gamma L^2}.
\label{d2}
\end{equation}
This is the well-known result for the mean first passage time from $0$
to $L$ for a freely diffusing particle with diffusion constant $D_{eff}$.
In this limit the flipping barrier 
behaves simply as an additional source of white noise of intensity
$\frac{\alpha^2}{2\gamma L^2}$. 
At the other extreme, when the flipping rate is
very low, $\lambda \rightarrow 0$, the barrier never flips as the process
moves from $0$ to $L$.  If the barrier is initially ``down'' ($\eta(0)=-1$),
then it remains down and the process is simply diffusively
driven toward the absorbing barrier by a constant force;
as $\alpha/D \rightarrow \infty$, the motion of the
system becomes increasingly ballistic. If the barrier is initially ``up''
($\eta(0)=1$), on the other hand, it remains up and the process moves between
the reflective barrier at $y=0$ and the absorbing barrier at $y=L$
{\em against}
a constant opposing force. The mean first passage time for such a process
grows exponentially with the barrier height 
as $D^2e^{\alpha/D}/\alpha^2$ \cite{Sesh}. In our
calculations either initial configuration is equally likely. The average
of these two possibilities is the leading term in Eq.~(\ref{r3}) (the
ballistic contribution is negligible):
\begin{equation}
\widetilde{T}_1 \rightarrow
\widetilde{T}_{static} =\frac{L^2D}{2\alpha^2} e^{\alpha/D}.
\label{d3}
\end{equation}
Between these two limits lies a regime
in which the mean first passage time is a minimum.  The minimum value
occurs at the resonant flipping rate whose leading term for large
values of $\alpha/D$ is
\begin{equation}
\gamma_{res} \sim \frac{\alpha}{\sqrt{2}L^2}
\label{d4}
\end{equation}
The leading contribution to the mean first passage time at this flipping
rate is
\begin{equation}
\widetilde{T}_{res} \sim (2+\sqrt{2})\frac{L^2}{\alpha}
\label{d5}
\end{equation}
and thus decreases with increasing $\alpha$. This result has the $L$
dependence of a diffusive process, but the effective diffusion coefficient
here is $\alpha$ and not $D$.  Note that the product
$\gamma_{res}\widetilde{T}_{res} = O(1)$.

\section{A simpler model showing resonant activation}
\label{simpler}

A surprising observation about the results of the Doering--Gadoua model
is that the resonant frequency $\gamma_{res}$ and the mean first passage
time at this resonant frequency, $\widetilde{T}_{res}$,
do not depend on the white noise intensity $D$ to leading order in $\alpha$.
This means that as
$\alpha/D \rightarrow \infty$,
the resonant properties become increasingly independent of the white noise
intensity. Indeed, the resonance therefore appears unaffected by and
unrelated to the white noise; in particular,
the minimum in the mean-first-passage time in the Doering--Gadoua model
appears not to arise from the coupling usually invoked
between the white noise and the random dynamics of the potential.

We have used somewhat equivocal language in this description because
the situation is in fact somewhat subtle.  The resonance
in the Doering-Gadoua model arises from two features: 1) the random
dynamics of the potential (that is, the random switching between barrier
up and barrier down) and, in particular, the initial average over
these dynamics, and 2) the nature of the reflecting boundary at $x=0$.
It is this latter feature, subtly influenced by the white noise,
that is especially noteworthy: the resonance
characteristics of the Doering--Gadoua result when $D\rightarrow 0$ are
{\em not} reproduced by simply setting $D=0$ to begin with in the model
equations.

To pursue this issue in more detail, let us consider the same
model equations as did Doering and Gadoua but now
in the absence of white noise from the outset.  In place of
Eq.~(\ref{basiceq}), the system thus evolves according to the
simpler Langevin equation
\begin{equation}
\dot{y}(\tau)=-g'(y)\eta(\tau).
\end{equation}
The solution of this mean first passage time problem is most easily
found by splitting $\widetilde{T}(y)$ into two components \cite{jaume1}:
$\widetilde{T}^+(y)$, the mean first passage time to $y=L$ when
$\eta(0)=+1$, and $\widetilde{T}^-(y)$, the mean first passage time to $y=L$
when $\eta(0)=-1$. The convenience of this representation lies in
the ease of expression of the boundary conditions in terms of 
$\widetilde{T}^{\pm}$. 

The boundary condition (\ref{con1}) is completely equivalent
to the so called ``immediate reinjection'' condition \cite{bala,jaume2}
\begin{equation}
	\widetilde{T}^+(y=0)=\widetilde{T}^-(y=0).
\end{equation}
In this case, whenever the system
reaches the boundary at $y=0$ the velocity immediately
changes its sign, that is, the driving noise $\eta(\tau)$
changes its value from $-1$ to $+1$. Note that here the boundary condition
directly affects the dynamics of the dichotomous barrier fluctuations
since arrival at the boundary
causes the noise to change its value.  It is easy to ascertain that
in terms of $\widetilde{T}(y)=(\widetilde{T}^+ + \widetilde{T}^-)/2$, the
``immediate reinjection" reflecting
boundary condition indeed translates to the Doering--Gadoua condition
(\ref{con1}), i.e.,
\begin{equation}
	\left.\frac{d\widetilde{T}(y)}{dy}\right|_{y=0}=0.
\end{equation} 
With this
boundary condition (together with the absorbing condition at $y=L$)
the mean first passage time from $y=0$ to $y=L$ if the initial
values $\eta(0)=\pm 1$ are equally probable is \cite{bala}
\begin{equation}
	\widetilde{T}_{1,ir} =\frac{\gamma L^4}{\alpha^2}+\frac{L^2}{\alpha}.
\end{equation}
Note that $\widetilde{T}_{1,ir}$ is a monotonically increasing function of
$\gamma$ and thus exhibits no resonance.  Clearly, this solution
is {\em not} the one approached by the Doering--Gadoua model when
$D$ is set to zero in the solution of the latter.  

There is another way to think of a reflecting boundary, namely, to  
assume that the boundary only limits the region of
movement of the system without interfering with the dynamics of the
dichotomous barrier fluctuations.
The dichotomous noise evolves according to its own dynamics, and changes its
value at random times that are independent of where the process $y(\tau)$
happens to
be.  Thus, if the system reaches the boundary $y=0$ when the noise
happens to be $\eta=-1$, the noise may retain this value according to
its own statistical properties. The process simply waits at the boundary,
until the noise switches to $\eta=1$ in the natural course of events.
We call this condition a ``natural'' reflecting boundary condition.
This behavior is implemented via the following boundary condition for
the mean first passage time components:
\begin{equation}
	\widetilde{T}^-(y=0)=\widetilde{T}^+(y=0)+\frac{1}{\gamma}
\end{equation} 
or, in terms of $\widetilde{T}(y)$, 
\begin{equation}
	\left.\frac{d\widetilde{T}(y)}{dy}\right|_{y=0}=-\frac{L}{\alpha}.
\end{equation} 
The solution for the mean first passage time is now
\begin{equation}
  \widetilde{T}_{1,n} =\frac{\gamma L^4}{\alpha^2}+\frac{2L^2}{\alpha}+
\frac{1}{2\gamma}.
\end{equation}
It is easily seen that $\widetilde{T}_{1,n}$ has a minimum at
$\gamma_{res}=\alpha/\sqrt{2}L^2$ (see Eq.~(\ref{d4})). 
In Fig.~\ref{fig4} we have plotted a realization of the process
$y(\tau)$ for the two
reflecting boundary conditions, the ``immediate reinjection" and
``natural". From this figure, it is clear that the two
boundary conditions lead to different results for the mean first passage
time.

The interesting point to note is that the $D\rightarrow 0$ limit 
of the mean first passage time in the Doering--Gadoua model 
is $\widetilde{T}_{1,n}$, that of the ``natural boundary,"
and {\em not} $\widetilde{T}_{1,ir}$, although the reflecting boundary condition
used for the solution of the Doering--Gadoua model is (\ref{con1}).
In the Doering--Gadoua model, no matter how weak the white noise, its
effects become dominant near the reflecting boundary. The white noise
allows reversal of the trajectory even infinitesimally close to the
reflecting boundary
without triggering a flip of the barrier, thus removing the dynamical
interference between the boundary and the dichotomous flipping process.
There is therefore a profound difference between the situation in which
white noise is present, albeit as weak as one wishes, and the situation
in which there is no white noise in the
first place --- the limit $D\rightarrow 0$ is discontinuous. 
In the Doering--Gadoua model there is a resonance in the mean first passage
time to absorption because, through the action of the white noise (no
matter how weak) the process {\em can} delay (increasingly as the white
noise intensity decreases) arrival at the
absorbing state if the barrier is up and does not flip of its own dynamics.
However, the characteristics of
the resonant flipping rate and the mean first passage time at resonance
do not explicitly depend on the intensity of the white noise.

In Fig.~\ref{fig5} we exhibit the two mean first passage times,
$\widetilde{T}_{1,ir}$ and $\widetilde{T}_{1,n}$.

\section{The Bier--Astumian model}
\label{bierastum}

The main conclusion that follows from the discussion of the preceding
sections is that the resonant effect in the ``toy" model of
Doering and Gadoua is
not of the same nature as the resonant activation in systems where the
activation process is exclusively due to the presence of white noise (i.e.,
nonzero temperature). In order to study the resonant process
in this latter situation, we return to the full model introduced by
Doering and Gadoua but now with $V_0>a$, so that there is always a barrier.
This problem was first analytically studied by Bier and
Astumian \cite{bier}. The approximation developed by these authors coincides
with the so called kinetic approximation introduced in Ref.~\cite{chris4}.
The main limitation of this method for the present purposes
is that it leads to a mean first passage time that does not 
exhibit a minimum. 

We have developed an approximation for the mean first passage time
to the absorbing boundary for high average barriers ($V_0 \gg 1$)
that does lead to a minimum and hence can be used to 
describe the resonance phenomenon analytically.  
We return to the Langevin equation (\ref{basiceq}) with
(\ref{white})--(\ref{g}).  Now, however, we take $\upsilon_0\pm\alpha \gg D$
or, in dimensionless quantities, $V_0\pm a \gg 1$.  The barrier thus
flips between two large values.

The general solution to Eq.(\ref{t1}) is
\begin{equation}
\label{t(x)}
T(x)=\frac{1}{V_0} (x-1)+A_1(e^{q_1x}-e^{q_1})+A_2(e^{q_2x}-e^{q_2})+
        A_3(e^{q_3x}-e^{q_3})
\end{equation}
where the coefficients $q_i$ are the three roots of the polynomial
equation
\begin{equation}
	q^3-2V_0q^2+(V_0^2-a^2- 2\lambda)q+2 \lambda V_0 =0
  \label{polynomial}
\end{equation}
and the constants $A_i$ have to be found from the boundary conditions
(\ref{con1})--(\ref{con4}). It can be demonstrated that
for $V_0> a$ the roots of
Eq.~(\ref{polynomial}) are all real, two of them positive and the other one
negative. The full expressions for the constants $A_i$ are
complicated and too long to be included here. However, it is possible to
derive shorter useful expressions for them as a series in $\lambda$.
In this case, the roots $q_i$ can be written as
\begin{eqnarray}
        q_1&=&-\frac{2 V_0}{V_+V_-}\lambda+
  \frac{4V_0(a^2+V_0^2)}{(V_+V_-)^3}\lambda^2+
  O(\lambda^3), \nonumber \\
q_2&=&V_-+\frac{\lambda}{V_-}-\frac{V_+}{2aV_-^3}\lambda^2+O(\lambda^3
),
  \nonumber \\
q_3&=&V_++\frac{\lambda}{V_+}+\frac{V_-}{2aV_+^3}\lambda^2+O(\lambda^3
),
  \label{roots}
\end{eqnarray}
where
\begin{equation}
V_{\pm}\equiv V_0 \pm a.
\end{equation}
When these expressions are introduced into Eq.~(\ref{t(x)}), the following
result is obtained for the mean first passage time to order $\lambda^2$:
\begin{equation}
        \label{t(x)aprox}
        T(x=0)=\frac{N_1 e^{V_-}+N_2 e^{V_+}+N_3 e^{2V_0}}
  {D_1+D_2 e^{V_-}+D_3 e^{V_+}},
\label{our}
\end{equation}
where the coefficients in the numerator are
\begin{eqnarray}
        N_1&=&V_+^2-\lambda\left(
  \frac{V_-V_+}{V_0}-\frac{V_0^2}{a V_-}-\frac{3aV_+}{V_-}
  +\frac{a^2 V_+}{V_0V_-}-\frac{a(V_0-5a)}{V_-^2}\right) \nonumber \\
        N_2 &=&V_-^2-\lambda\left(
  \frac{V_-V_+}{V_0}+\frac{V_0^2}{a V_+}+\frac{3aV_-}{V_+}
  +\frac{a^2 V_-}{V_0V_+}+\frac{a(V_0+5a)}{V_+^2}\right) \nonumber\\
         N_3 &=&4\lambda \left\{1+\lambda\left[
  \frac{(2 V_0-1)}{V_+ V_-}-\frac{4a^2}{(V_+ V_-)^2}\right]\right\},
\label{abc}
\end{eqnarray}
and those of the denominator are
\begin{eqnarray}
        D_1&=&2(V_+V_-)^2+2\lambda (a^2+3V_0-2V_0 V_+V_-)\nonumber \\
  D_2&=&2\lambda\left[V_+^2+
  \frac{\lambda}{aV_-^2}\left(
  a V_+^2 V_--V_+ V_-^2+2a^3-6 a^2 V_0\right)\right] \nonumber \\
  D_3&=&2\lambda\left[V_-^2+
  \frac{\lambda}{aV_+^2}\left(
  a V_-^2 V_++V_- V_+^2+2a^3+6 a^2 V_0\right)\right].
        \label{delta}
\end{eqnarray}

This approximation is one order higher in $\lambda$ than the one derived
in Ref.~\cite{bier}, which is equivalent to the so-called kinetic
approximation \cite{chris4}.
The advantage of our new approximation 
is that it shows a minimum as function of the frequency
$\lambda$.  When $V_0\gg 1$ the resonant frequency can be
calculated explicitly:
\begin{equation}
  \lambda_{res} \sim
  \left[\frac{a(e^{2a}-1)^2V_0^3}{4 e^{2a}(1+a-e^{2a}+ae^{2a})}
  \right]^{1/2} \exp\left(-\frac{V_-}{2}\right)
\label{explicit}
\end{equation}
and the associated minimal mean first passage time reads
\begin{equation}
  T_{res} \sim \frac{2}{V_-^2+e^{-2a}V_+^2}\; e^{V_-}.
\label{done}
\end{equation}

The analytic expressions (\ref{our})--(\ref{done}) are the principal
results of this paper.

As we did in the Doering--Gadoua case, it is useful to exhibit explicitly
various limiting results in the original units so that the dependence on
system parameters is clarified. 

When the flipping rate is large ($\lambda \gg a^2$ or
$\gamma \gg \alpha^2/DL^2$) the result (\ref{our}) reduces to
\begin{equation}
T_1\rightarrow T_{white}=\frac{2+a^2/\lambda}{2V_0^2} \exp\left(\frac
{2 V_0}{2+\frac{a^2}{\lambda}}\right).
\label{wn}
\end{equation}
or, in the original units,
\begin{equation}
\widetilde{T}_1 \rightarrow \widetilde{T}_{white}
= \frac{L^2D_{eff}}{\upsilon_0^2}
e^{\upsilon_0/D_{eff}}
\label{wn2}
\end{equation}
where $D_{eff}$ is the effective diffusion coefficient defined in
Eq.~(\ref{d2}). This is the appropriate and familiar result for 
activation over a barrier of height $\upsilon_0$ with diffusion coefficient
$D_{eff}$. At the other extreme, as $\lambda$ becomes small,
the kinetic approximation \cite{bier,chris4}
is valid and the mean first passage time (\ref{our}) reduces to
\begin{equation}
        T_1\rightarrow T_{kin}=\frac{2 \lambda+(k_+ + k_-)/2}
                {k_+ k_- +\lambda(k_+ + k_-)},
\label{kinetic}
\end{equation}
where
\begin{equation}
k_{\pm}=V_{\pm}^2 e^{-V_{\pm}}.
\label{rate}
\end{equation}
If $\lambda$ becomes so small that the time scale of barrier
fluctuations is much slower than the escape time, then this further
simplifies to
\begin{equation}
T_1\rightarrow T_{kin} \sim \frac{1}{2}\left( \frac{1}{k_+} + \frac {1}{k_+}
\right)
\end{equation}
which is just the arithmetic mean associated with the two possible initial
barrier configurations (see discussion surrounding Eq.~(\ref{d3})). 
In the original units
\begin{equation}
\widetilde{T}_1 \rightarrow \widetilde{T}_{kin} =
\frac{L^2D}{2\upsilon_+^2} e^{\upsilon_+/D} +
\frac{L^2D}{2\upsilon_-^2} e^{\upsilon_-/D}.
\label{dd3}
\end{equation}

Between these two limits lies the resonance regime where the mean first
passage time is shorter than either the ``white" or the ``static" results.
In the original units the mean first passage time at resonance,
Eq.~(\ref{done}), reads
\begin{eqnarray}
\widetilde{T}_{res} &\sim& \frac{2L^2D}{(\upsilon_-^2+e^{-2a}\upsilon_+^2)}
e^{\upsilon_-/D}\nonumber\\
&&\nonumber\\
&\sim& \frac{2L^2D}{\upsilon_-^2} e^{\upsilon_-/D}
\label{dd4}
\end{eqnarray}
where the second line, valid if $a\geq 1$, serves to stress the point that
the resonant mean first passage time is essentially the usual passage
time over the lower of the two barriers.  It is not particularly
instructive to exhibit the full expression (\ref{explicit}) for the
resonance frequency in the original units, but if $a\geq 1$ we can display
the shorter expression
\begin{equation}
\gamma_{res}\sim \frac{\upsilon_0^{3/2}}{2L^2D^{1/2}}e^{-\upsilon_-/2D}.
\label{dd5}
\end{equation}
It should be noted that both the resonant mean first passage and the
resonant frequency depend on the intensity of the white noise, as does
their product. This dependence appears in the exponents as well as
prefactors.

A general feature of our solution and, more generally, of the resonant
activation phenomenon is that with increasing barrier height
the resonance phenomenon becomes less and less sharp: a long flat region
develops around the resonant frequency, a fact that has been
explicitly noted in earlier work \cite{reim3}. Analysis of 
Eq.(\ref{t(x)aprox}) makes it possible to estimate analytic bounds
of this flat region, which spans the range 
\begin{equation}
\frac{V_+^2e^{-2a}+V_-^2}{4} \;e^{-V_-} \ll \lambda \ll
\frac{V_+V_-}{2V_0}.
\end{equation}
Thus, rather than stressing the resonance aspect of the problem,
it might be more accurate to describe the time scale of the
activation process as relatively insensitive to the parameters of the
system except in the limits of very low and very high barrier fluctuation
rates.  As noted above,
if the barrier fluctuations are sufficiently slow, then an initially
high barrier remains that way essentially forever and the system on average
crosses it before the barrier flips.  Passage over the higher barrier then
dominates the mean first passage time.  At the other extreme, when
the barrier fluctuations are very rapid, crossing occurs essentially over
the average barrier.   However, over most parameter ranges the mean
first passage time is
essentially determined by passage over the lower barrier -- the system
can avoid passage over the higher barrier by ``waiting" for it to flip.
Provided the waiting time is shorter than the time it would take the
system to cross the high barrier, flipping will occur first and the
system will cross when the barrier is lower (unless flipping is too rapid). 
This process is
most efficient (but not dramatically more efficient -- hence the flat
behavior) at the resonance frequency.

In Fig.~\ref{fig6} the mean first passage time and the different
approximations explained above have been
plotted for $V_0=11$ and $a=1$.  Our approximation clearly captures the
resonance behavior extremely accurately and for that matter the behavior
of the mean first passage time over a broad range of barrier fluctuation
rates. Fig.~\ref{fig7} again shows the mean first passage time, but
now with $V_0=15$ and $a=1$.  This figure serves to reconfirm the
agreement of our results with the numerical ones, and also 
illustrates the flattening of the region around the resonance 
as the height of the barrier increases.
Fig.~\ref{fig8} shows the
resonance frequency as a function of the barrier height.  Again, our
approximation clearly captures the exact results extremely well for
sufficiently high barriers.

\section{Activation driven by a square wave function}
\label{square}

It is interesting to explore whether the
resonant activation phenomenon requires that the barrier
fluctuate stochastically, or whether it also occurs when a noisy
process crosses a barrier that changes periodically.  Indeed, stochastic
fluctuation of the barrier is not a requirement.

To investigate the activation process when the barrier oscillates
periodically between a higher value and a lower value, we replace the
dichotomous noise $\eta(\tau)$ in Eq.~(\ref{basiceq}) with a square wave
$w(\tau)$, a periodic function that alternately takes on the values $+1$
and $-1$. The changes from one to the other occur
at a constant frequency $\gamma$. The period of the square
wave function is thus $2/\gamma$.  

The Fokker-Planck equation describing the evolution of the probability for
the system now includes a time-periodic potential. The problem can be
tackled analytically using Floquet theory.  We simply state qualitatively
the results that one obtains with this exact approach, but then follow
a simpler approach to arrive at some quantitative conclusions.

We continue our discussion in terms of dimensionless variables and
parameters.
Exact solution of the problem does, as noted above, also lead to
resonant activation when the barrier changes from higher to lower
periodically, with a resonance behavior very similar to that of the
stochastic case.  In other words, the mean first passage time is large
when the period of oscillation is very slow and also when it is very fast.
As before, and for the same physical reasons, in the former case the
mean first passage time is dominated
by the high barrier $V_+$ and in the latter case it is determined
by the average barrier $V_0$. Again as before, between these two
limits there is a flat region (i.e., rather insensitive to the
parameter values) where the mean first passage time is
determined primarily by the lower barrier $V_-$.  The only difference
between this problem and the stochastic one lies in the detailed way in
which the mean first passage time changes from one behavior to the other.

To find the mean first passage time at the slow-barrier-modulation end of the
problem (where the difference between stochastic and periodic modulation
is most pronounced), we recall that
for a fixed barrier of height $V$ the probability that the process has {\em
not} yet crossed the barrier at time $t$ (i.e., the survival probability at
time $t$), is exponential \cite{katja2}, $e^{-kt}$, where
the crossing rate $k=V^2e^{-V}$ [cf. Eq.~(\ref{rate})]. If
the barrier is not fixed, but instead changes slowly from one value to the
other at predetermined periodic time intervals $\Delta \equiv\lambda^{-1}$
with $\lambda\lesssim 1$,
we can track the trajectory of the system explicitly and write down
an expression for the survival probability $S(t)$ that the
process has not yet crossed the barrier at time $t$.
If initially the barrier is $V_+$ and the time $t=0$ corresponds to the
beginning of a period, then 
\begin{eqnarray}
  S^+=\mbox{Prob}\{T^+ >t\}=q_+^n q_-^n e^{-k_+(t-2n\Delta)},\qquad && 
		2n\Delta < t \le (2n+1)\Delta; \nonumber \\
  S^+=\mbox{Prob}\{T^+ >t\}=q_+^{n+1} q_-^n e^{-k_-[t-(2n+1)\Delta]},\qquad &&
		(2n+1)\Delta < t \le (2n+2)\Delta
\end{eqnarray} 
where $n=0,1,2,\ldots$, and $q_+$ and
$q_-$ are the probabilities that a
crossing event does not happen when the barrier is, respectively, $V_+$ and
$V_-$. The assumption about the statistics of the crossing events yields 
\begin{equation}
  q^{\pm}=e^{-k_{\pm}/\lambda}.
\end{equation}
The mean first passage time $T^+$ can then be calculated directly as a
moment of this probability.
The survival probability $S^-$ and associated mean first passage time
$T^-$ when the barrier is initially $V_-$ is similarly
obtained.
To compare most directly with the stochastic results we assume
that initially the barrier is equally likely to be $V_+$ or $V_-$. 
A short calculation then leads to the following result for the
mean first passage time when $\lambda \lesssim 1$:
\begin{equation}
	T=\frac{T^++T^-}{2}=
	\frac 12 \left(\frac{1}{k_+}+\frac{1}{k_-}\right)
	+\frac 12  \left(\frac{1}{k_+}-\frac{1}{k_-}\right)
	\frac{q_+-q_-}{1-q_+q_-}.
\label{squareeq}
\end{equation}

This result corresponds to the same level of approximation as the kinetic
result (\ref{kinetic}).   At very low frequencies, $\lambda \rightarrow 0$,
the mean escape time is correctly given by an average of the escape time
$1/k_+$ when the barrier is $V_+$ and $1/k_-$
when the barrier is $V_-$.  This kinetic approximation also does not
exhibit a minimum because it does not behave correctly 
when $\lambda\rightarrow \infty$; instead, it converges to the same
value as the kinetic approximation (\ref{kinetic}), that is,
to $(k_++k_-)/2$. 

In Fig.~\ref{fig9} the approximation (\ref{squareeq})
to the mean escape time for the
activation process driven by a periodic signal is compared with the
escape time for the same system driven by dichotomous noise.  
The difference between the two is noticeable in the decrease of
the mean first passage time with increasing flipping rate --- the
dependence on flipping rate is considerably sharper in the periodic
case than in the random case. 
A similar effect has been observed recently in systems that exhibit
coherent stochastic resonance \cite{pep}.  The
minimum first passage time and resonance flipping rate are
essentially identical in the two cases.

\section{Conclusion}
\label{conclusions}

We have revisited the problem of resonant activation, that is, of the
mean escape time of a particle driven by white noise of intensity $D$
over a barrier of randomly fluctuating height.  The initial position
of the particle is $y=0$ and the barrier is at $y=L$.
A substantial recent literature \cite{DG,bier,DN,OU,marchi,chris4} deals with
this problem, but the results to be found in the literature
are almost exclusively numerical.  
The distribution of barrier fluctuations is
typically taken to be either dichotomous (i.e., the barrier
fluctuates between two values) or Gaussian.  The correlation
function of the barrier fluctuations is usually assumed to be exponential
and thus characterized by a rate parameter $\gamma$.
The quantity of interest is the mean escape time $\widetilde{T}_1$
of the particle over the barrier as a function of $\gamma$. It is
observed that $\widetilde{T}_1$ vs. $\gamma$ exhibits a minimum, i.e., there is an
optimal barrier fluctuation rate that minimizes the escape time of the
particle.  This minimum defines the resonant activation phenomenon.

In this paper we have concentrated on dichotomous fluctuations and on
triangular potential barriers, so our quantitative results are restricted
to these cases.  However, we believe that our results provide insights
beyond these specific conditions.  In particular, they provide insights
for barriers whose fluctuations are bounded between an upper value
$\upsilon_+$ and a lower value $\upsilon_-$.  The applicability of our
conclusions to Gaussian fluctuations is therefore less certain, but
below we will present some conjectures for this case as well.

A variety of approaches to the problem of the escape over a fluctuating
barrier of bounded variation provide excellent and consistent
analytic approximations to the escape rate 
in the limiting cases of very slow
barrier fluctuations and of very fast barrier fluctuations.  
In the slow fluctuation case, the so-called ``kinetic approximation"
\cite{chris4} captures the behavior of the system very well. 
In the limit of very slow fluctuations ($\gamma \rightarrow 0$)
the barrier retains its initial height throughout the process. The mean first
passage time for the ensemble is then just the mean first passage time
averaged over the initial distribution of barrier heights.  For example,
in the dichotomous case if the height of the high barrier is
$\upsilon_+$ then the mean escape time
as $\gamma \rightarrow 0$ is determined by the mean escape time over this
high barrier (the mean escape time over the lower barrier being
negligible in comparison):
\begin{equation}
\widetilde{T}_{static} \sim
\frac{L^2D}{2\upsilon_+^2} e^{\upsilon_+/D}.
\label{c1}
\end{equation}
This is the result captured, for instance, in Eq.~(\ref{d3}) and in
Eq.~(\ref{dd3}.  Clearly, this result is determined in part by the 
assumption (generally made in the literature) that
an initial average over an ensemble of barrier heights is appropriate.
The entire discussion that follows, including the occurrence of a
resonance, is dependent on such an initial average or at least on the
assumption that a finite fraction of realizations begin with a barrier
configuration that is higher than the lowest barrier.

At the opposite extreme, when the barrier fluctuations are very rapid
($\gamma \rightarrow \infty$), the main effect of the flipping barrier
is to increase the effective intensity of the white noise.  The
escape then occurs over the average barrier, with a diffusion
coefficient $D_{eff}$ which exceeds $D$ by an amount determined by
the detailed
distribution of barrier fluctuations.  If the average barrier height is
$\upsilon_0>0$ then the mean escape time in this limit is
\begin{equation}
\widetilde{T}_{white} \sim
\frac{L^2D_{eff}}{\upsilon_0^2} e^{\upsilon_0/D_{eff}}. 
\label{c2}
\end{equation}
If the average barrier height is zero, then
\begin{equation}
\widetilde{T}_{white} = \frac{L^2}{2D_{eff}}.
\label{c3}
\end{equation}
These are the results captured in Eqs.~(\ref{d1}) and (\ref{wn2}).
In any case, the escape time is clearly smaller in the fast barrier
fluctuation limit than in the slow barrier fluctuation limit.

In the literature, each of the above approximations had been carried
sufficiently far to deduce the behavior of the escape time as one moves
away from the strict limits.  Thus, within the kinetic approximation
it can be shown that the escape time decreases with increasing $\gamma$. 
At the opposite limit, it can be shown that the escape time decreases with
decreasing $\gamma$.  These two results clearly point to a minimum for
some finite value of $\gamma$, but neither approximation is sufficient to
actually capture the minimum.  Our goal here has been to develop a
single approximation to capture this minimum, and in this we succeeded.

However, we found in the process that a distinction needs to be made
between two cases that lead to a different physical origin and parameter
dependences for the resonant flipping rate and the associated escape
time.  These two cases have in the literature been treated more or less 
as one because both involve dichotomous fluctuations, but they are
in fact very different.  One of these is the case in which the
``barrier" fluctuates between an ``up" or positive (barrier)
configuration of height $\upsilon_+=\alpha$ and a
``down" or negative (valley) configuration of height
$\upsilon_-=-\alpha$.
We have called this the Doering--Gadoua model \cite{DG}.
In the other case, the
barrier fluctuates between a high value 
$\upsilon_+=\upsilon_0+\alpha$ and a lower (but
still positive) value $\upsilon_-=\upsilon_0-\alpha$. 
We have called this the Bier--Astumian model \cite{bier}. 
For each model we found a
single expression for the mean first passage time that has a minimum,
and we compared our results with exact ones obtained numerically.  The
agreement in both cases is excellent for almost the entire range
of flipping rates, and in particular over a broad range surrounding the
resonance.

The distinctive aspect of the Doering--Gadoua model is the fact that
part of the time the ``barrier" is really a valley, so that the particle
can essentially roll rather than climb towards $L$ during these times.
We found an explicit expression for the resonant flipping rate and the
resonant mean first passage time in this case:
\begin{equation}
\gamma_{res} \sim \frac{\alpha}{\sqrt{2}L},
\end{equation}
\begin{equation}
\widetilde{T}_{res} \sim (2+\sqrt{2})\frac{L^2}{\alpha}.
\end{equation}
The noteworthy fact about these results is that neither the resonant
flipping rate nor the resonant escape time depend explicitly on the
intensity $D$ of the white noise.  This fact seems not to have been noted
before.  We then went on to explore whether in fact this resonance is
observed in a process defined by the Doering--Gadoua model
with no white noise from the outset, and found that there is no resonance
for such a model!  We explained this apparent contradiction by noting a
discontinuity in the $D\rightarrow 0$ limit of the problem and by
presenting a modified set of boundary conditions that does lead to
a resonance (precisely the Doering--Gadoua resonance) in the absence
of white noise.  

In order to obtain a result for the escape time in the Bier--Astumian
model that captures the resonant behavior we found that we had to
retain terms in our solutions to one power higher in $\gamma$ than
had been done previously (the lower orders yielded only the kinetic
approximation) \cite{bier}.  With this, we identified the resonant
frequency and escape times as 
\begin{equation}
\gamma_{res}\sim \frac{\upsilon_0^{3/2}}{2L^2D^{1/2}}e^{-\upsilon_-/2D}
\end{equation}
and (more complete results are found in Sec.~\ref{bierastum})
\begin{equation}
\widetilde{T}_{res}
\sim \frac{2L^2D}{\upsilon_-^2} e^{\upsilon_-/D}.
\end{equation}
We noted that that in this case both the resonance frequency and the
mean first passage time at the resonance frequency depend on the intensity
of the white noise, as does their product.  At resonance
the escape over the barrier occurs primarily when the barrier is at its
lowest. We also noted that the dependence on the flipping rate,
especially for high barriers, is very flat: there is a broad range of
flipping rates where passage over the barrier occurs primarily when
the barrier is low. In this broad range of flipping rates the
escape time over the high
barrier is so long that the barrier is likely to flip to its lower height
before the escape is completed.

We also discussed the fact that the resonant activation phenomenon does
not require a fluctuating barrier --- it also occurs if the barrier
oscillates periodically between the high and low values.  The behavior
of the escape time at low and high
oscillation periods is the same as in the dichotomous fluctuation case,
and at intermediate oscillation periods a resonance effect is also observed.  

Finally, we note that our analysis does not address the case of Gaussian
barrier fluctuations, that is, of Ornstein--Uhlenbeck barrier
fluctuations \cite{marchi}.  The results for such barrier fluctuations
with fixed variance \cite{iwan} should be similar to our results for
dichotomous noise. In particular, the escape time for the model analogous
to that of Doering and Gadoua ($V_0=0$) will show a minimum even in the
absence of white noise.

\section*{Acknowledgments}
We gratefully acknowledge partial support of this research by the CICYT
No. PB96-0188.  One of us (K. L.) 
gratefully acknowledges the partial support of this research by the
U. S. Department of Energy through Grant No. DE-FG03-86ER13606.

\appendix
\setcounter{equation}{0}
\renewcommand{\theequation}{\thesection\arabic{equation}}
\section{Survival probability and first passage time moments}
The moments of the first passage time from $x=0$ to $x=1$ can be
obtained
from the survival probability $S(x,t)$ that the system evolving
according
to Eq.~(\ref{basiceq}) (appropriately scaled to dimensionless variables)
with a reflecting boundary at $x=0$ and an
absorbing
boundary at $x=1$ has not left the interval $(0,1)$ at time t. This
survival probability obeys the following
partial differential equation (a detailed derivation and original
references for the survival probability in an interval terminated by
two absorbing boundaries are presented in Ref.~\cite{Porra}):
\begin{equation}
 {\cal L}^2 S+2 \lambda {\cal L} S =
 a^2 \frac{{\partial}^2 S}{\partial x^2},
\label{2.1}
\end{equation}
where ${\cal L}$ is the differential operator
\begin{equation}
 {\cal L}\equiv\frac{\partial}{\partial t} +
 V_0 \frac{\partial}{\partial x} -
\frac{{\partial}^2}{\partial x^2}.
\end{equation}
Note that Eq.~(\ref{2.1}) is a second-order partial differential
equation in the time variable and a fourth order partial differential
equation in the state variable.
Therefore, two initial conditions and four boundary conditions are
needed to solve it. The initial conditions are
\begin{eqnarray}
 S(x,0)&=&1,\\ &&\nonumber \\
        \left. \frac{\partial S}{\partial t} \right|_{t=0} &=&0.
\end{eqnarray}
For the absorbing trap at $x=1$, the boundary conditions read
\begin{eqnarray}
 &&S(1,t)=0, \\ &&\nonumber \\
        &&\left. \left( {\cal L} +V_0
                \frac{\partial}{\partial x} 
	-V_0^2+a^2
		\right) \frac{\partial S}{\partial x}
                \right|_{x=1}=-V_0 \delta(t),
\end{eqnarray}
and for the reflecting boundary at $x=0$ they are
\begin{eqnarray}
        &&\left. \frac {\partial S}{\partial x} \right|_{x=0}=0, \\
        &&\nonumber \\
        &&\left( \frac{\partial}{\partial t} -
                \frac{\partial^2}{\partial x^2}\right)
                \left. S\right|_{x=0}=0.
\end{eqnarray}

The first-passage-time moments $T_n$ are related to the survival
probability according to
\begin{equation}
T_n(x)=n\int_0^\infty t^{n-1}S(x,t)dt.
\end{equation}
Clearly $T_0(x)=1$ by normalization. $T_1(x)$ is the mean first
passage
time to $1$ for a process that starts at $X(0)=x$; $T_2(x)$ is the
second
moment of the distribution, so that the variance of the distribution of
mean first passage times is $\sigma^2 \equiv T_2 - T_1^2$.

Equations for the first passage time moments can be obtained
by multiplying Eq.~(\ref{2.1}) by $t^{n-1}$ and integrating over time by
parts.  The following recursive-differential equation is easily found:
\begin{equation}
        {\cal L }_D^2 T_n-2 \lambda  {\cal L}_D T_n-
        a^2 \frac{ d^2 T_n}{d x^2}=g_n,
\label{recursive}
\end{equation}
where
\begin{equation}
	g_n \equiv n(2 \lambda -2 {\cal L}_D)T_{n-1}-n(n-1)T_{n-2},
\end{equation}
and $ { \cal L}_D $ is the Fokker-Planck operator
\begin{equation}
		{ \cal L}_D \equiv -V_0 \frac{d}{d x}
		+ \frac{d^2}{d x^2}
\end{equation}
with $T_{-1}\equiv 0$.  The boundary conditions can be obtained directly
from those of the survival probability:
\begin{eqnarray}
	{\mbox{1.  }}& &T_n(1)=0 \\ &&\nonumber \\
	{\mbox{2.  }}& &\left({\cal L}_D
	-V_0 \frac{d}{d x} + V_0^2-a^2
	\right) \left. \frac{d T_n}{d x} \right|_{x=1}
	=- n \left. \frac{d T_{n-1}}{dx} \right|_{x=1}+V_0 \delta_{n,1} \\
	&&\nonumber\\
	{\mbox{3.  }}& &\left. \frac{d T_n}{d x} \right|_{x=0}=0 \\
	 &&\nonumber\\
	{\mbox{4.  }}& & 
	\left. \frac{d^2 T_n}{d x^2} \right|_{x=0}
				=- n T_{n-1}(0).
\label{bc}
\label{boundary}
\end{eqnarray}

\newpage

\begin{figure}
\caption
{Schematic of the fluctuating potential barrier problem.}
\label{fig1}
\end{figure}

\begin{figure}
\caption
{Mean first passage time as a function of barrier fluctuation rate
for the Doering--Gadoua model.  The barrier fluctuates between the
up and down positions with slopes $a$ and $-a$ respectively.
Solid curve with circles: exact mean first passage time
obtained numerically.  Solid curve with squares:
our analytic result (\ref{r1}). Dotted curve:
the low--frequency approximation (\ref{r3}). The parameter $a=8$.}
\label{fig2}
\end{figure}

\begin{figure}
\caption
{Second moment of the first passage time distribution for
the Doering--Gadoua model as a function of
barrier fluctuation rate, with $a=8$.  
Solid curve with circles: exact second moment obtained numerically.
Solid curve with squares: our analytic result (\ref{t2}).}
\label{fig3}
\end{figure}

\begin{figure}
\caption
{Typical trajectories near the reflecting boundary for the two types
of boundaries discussed in the text. Dashed lines: ``immediate rejection"
boundary. Solid lines: ``natural" boundary.}
\label{fig4}
\end{figure}

\begin{figure}
\caption
{Mean first passage time to $L=1$ as a function of barrier fluctuation rate
for the two types of reflecting boundaries discussed in the text.
Dashed curve with squares: ``immediate rejection"
boundary. Solid curve with circles: ``natural" boundary.}
\label{fig5}
\end{figure}

\begin{figure}
\caption
{Mean first passage time as a function of barrier fluctuation rate
for the Bier--Astumian model.  The barrier fluctuates between the high
value $V_0+a$ and the lower value $V_0-a$, with $V_0=11$ and $a=1$.
Solid curve with circles: exact mean first passage time
obtained numerically. Dotted curve: kinetic approximation 
(\ref{kinetic}). Dashed curve: white noise approximation 
(\ref{wn}). Solid curve with squares: our result (\ref{our}).}
\label{fig6}
\end{figure}

\begin{figure}
\caption
{Mean first passage time as a function of barrier fluctuation rate
for the Bier--Astumian model with $V_0=15$ and $a=1$. 
Solid curve with circles: exact
mean first passage time obtained numerically. Dashed curve: white
noise approximation (\ref{wn}). Solid curve with squares:
our result (\ref{our}).}
\label{fig7}
\end{figure}

\begin{figure}
\caption
{Resonance frequency as a function of average barrier height for
the Bier--Astumian model with $a=1$.
Circles: exact numerical results. Solid curve: our result
(\ref{explicit}).}
\label{fig8}
\end{figure}

\begin{figure}
\caption
{Mean first passage time as a function of barrier oscillation/fluctuation
rate.  Square symbols: kinetic approximation for square wave barrier
oscillations.  Circles: fluctuating barrier.  Barrier parameter values:
$V_0=11$ and $a=1$.}
\label{fig9}
\end{figure}

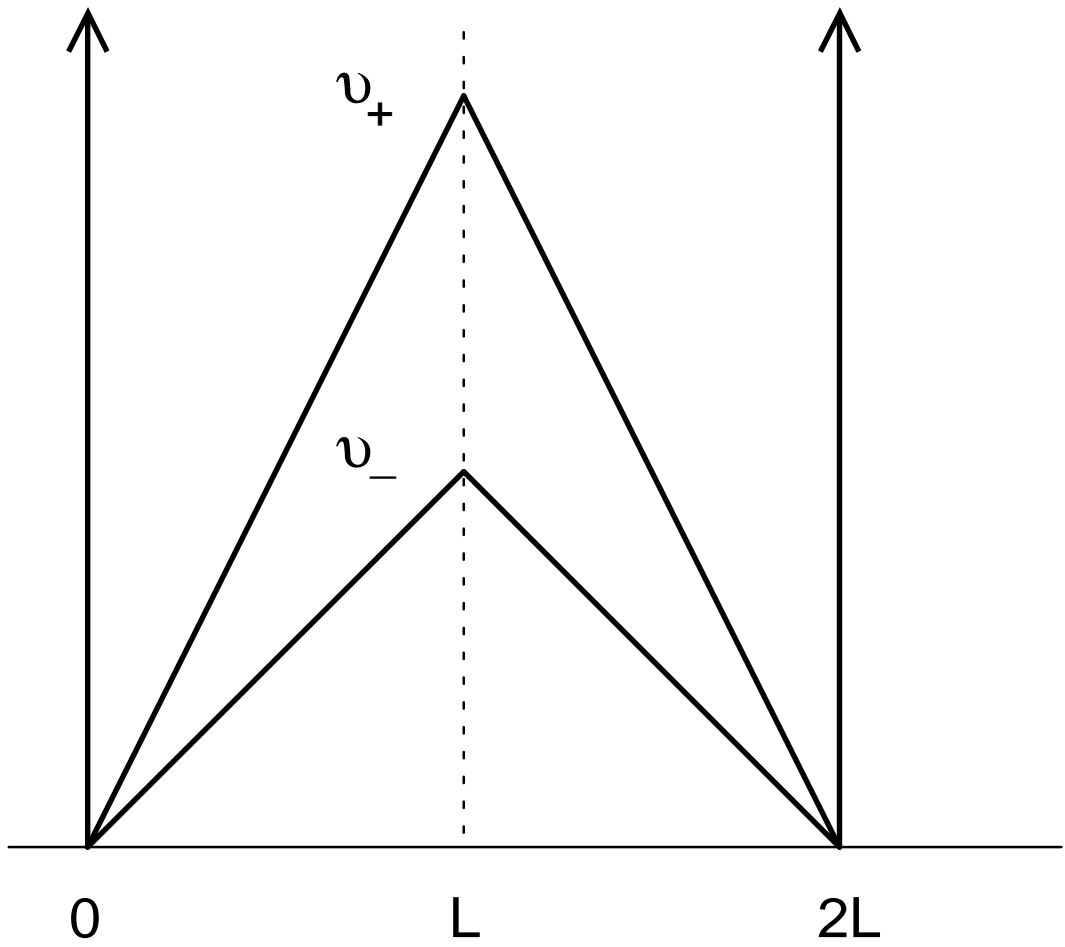
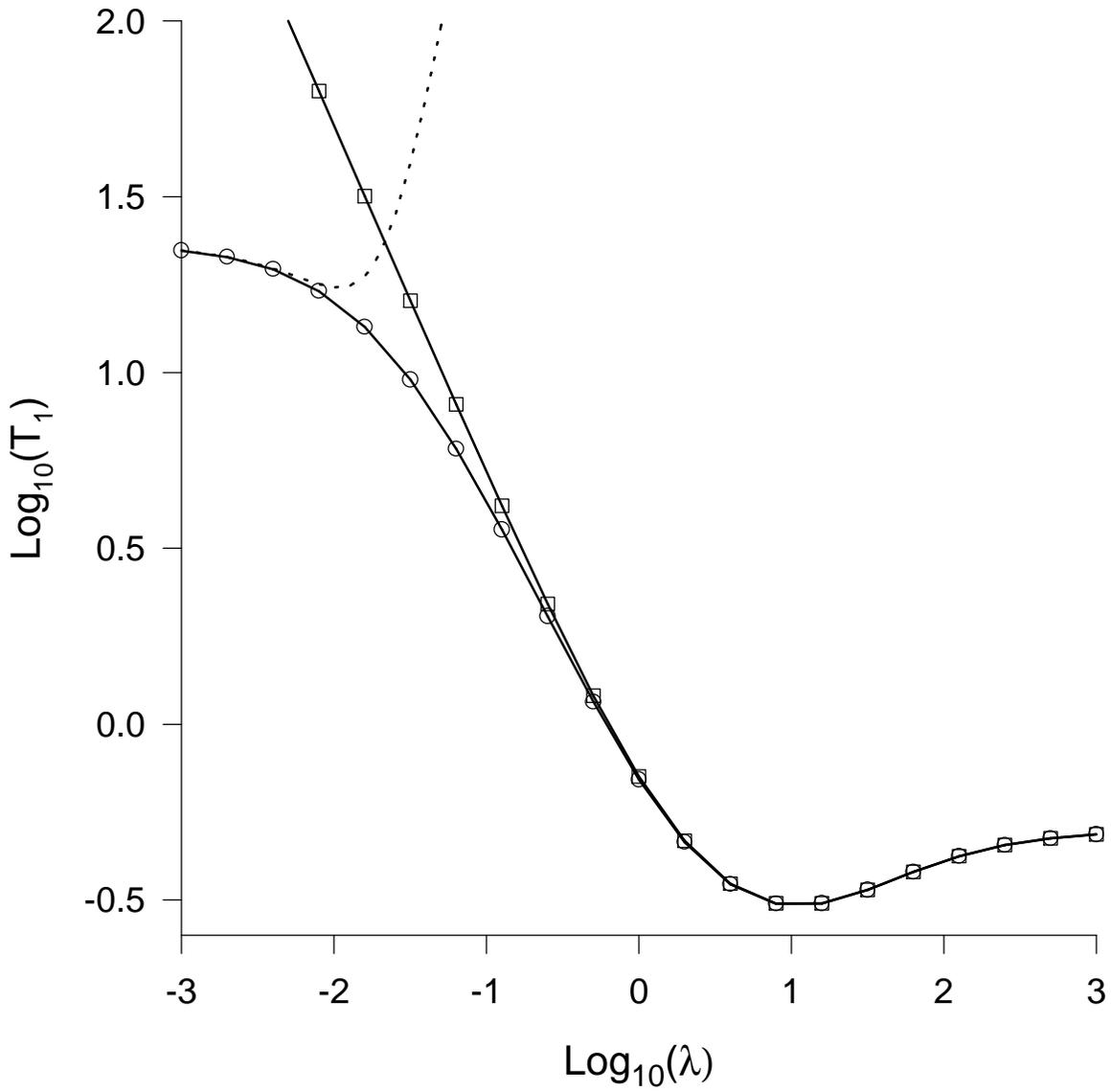
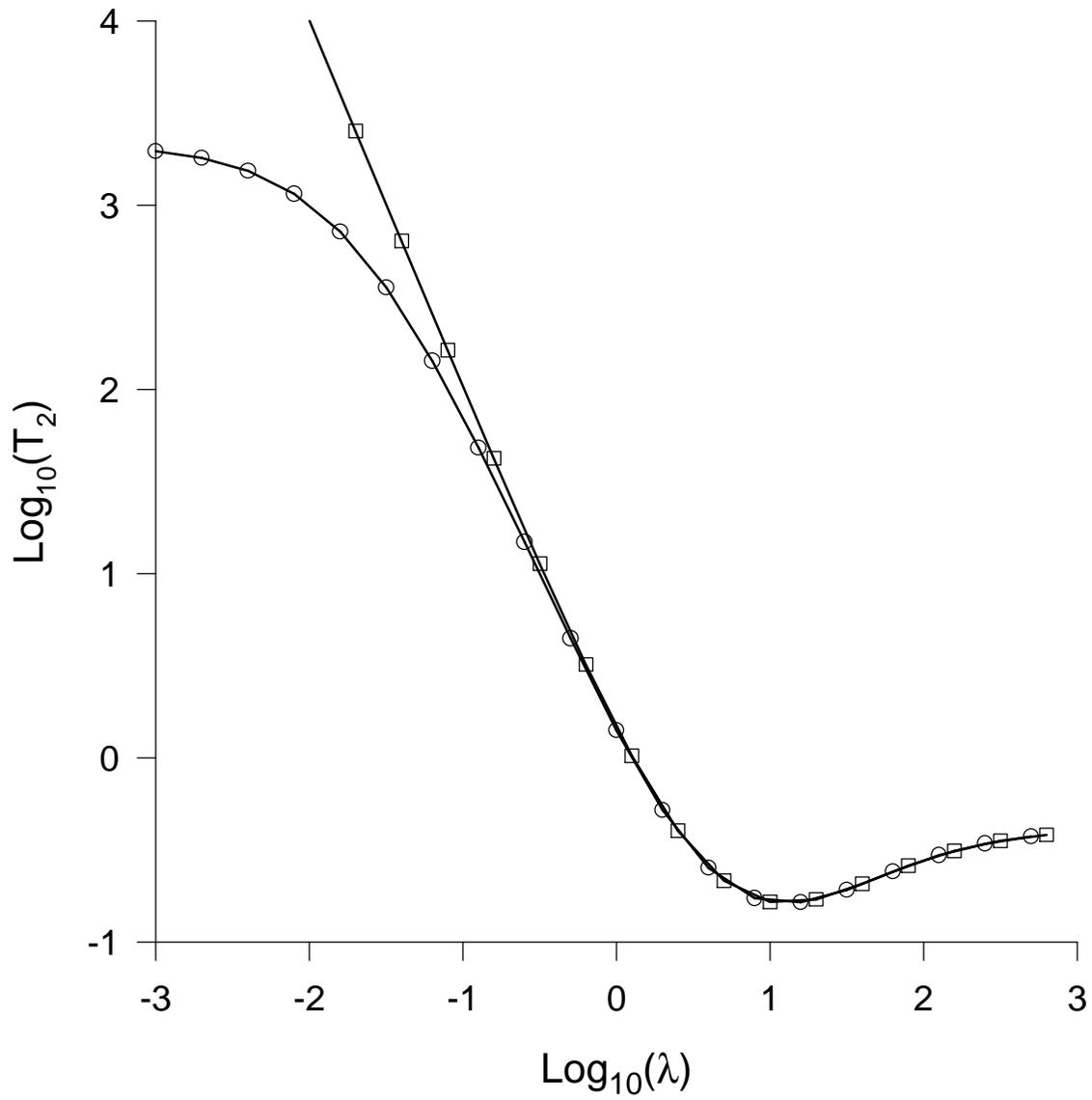
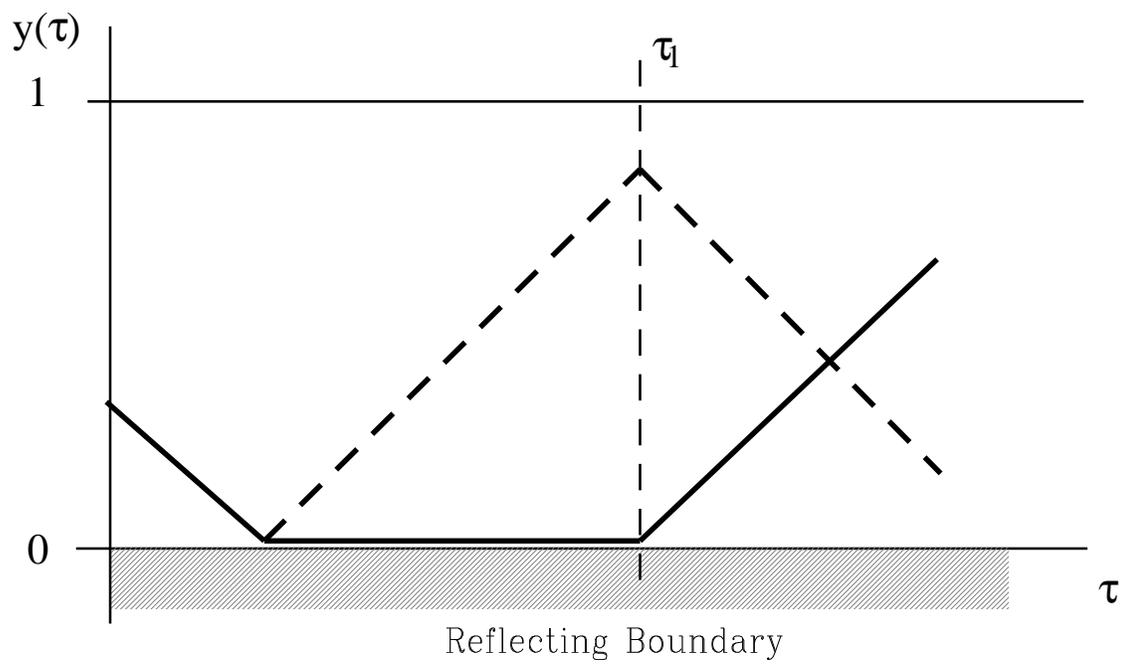
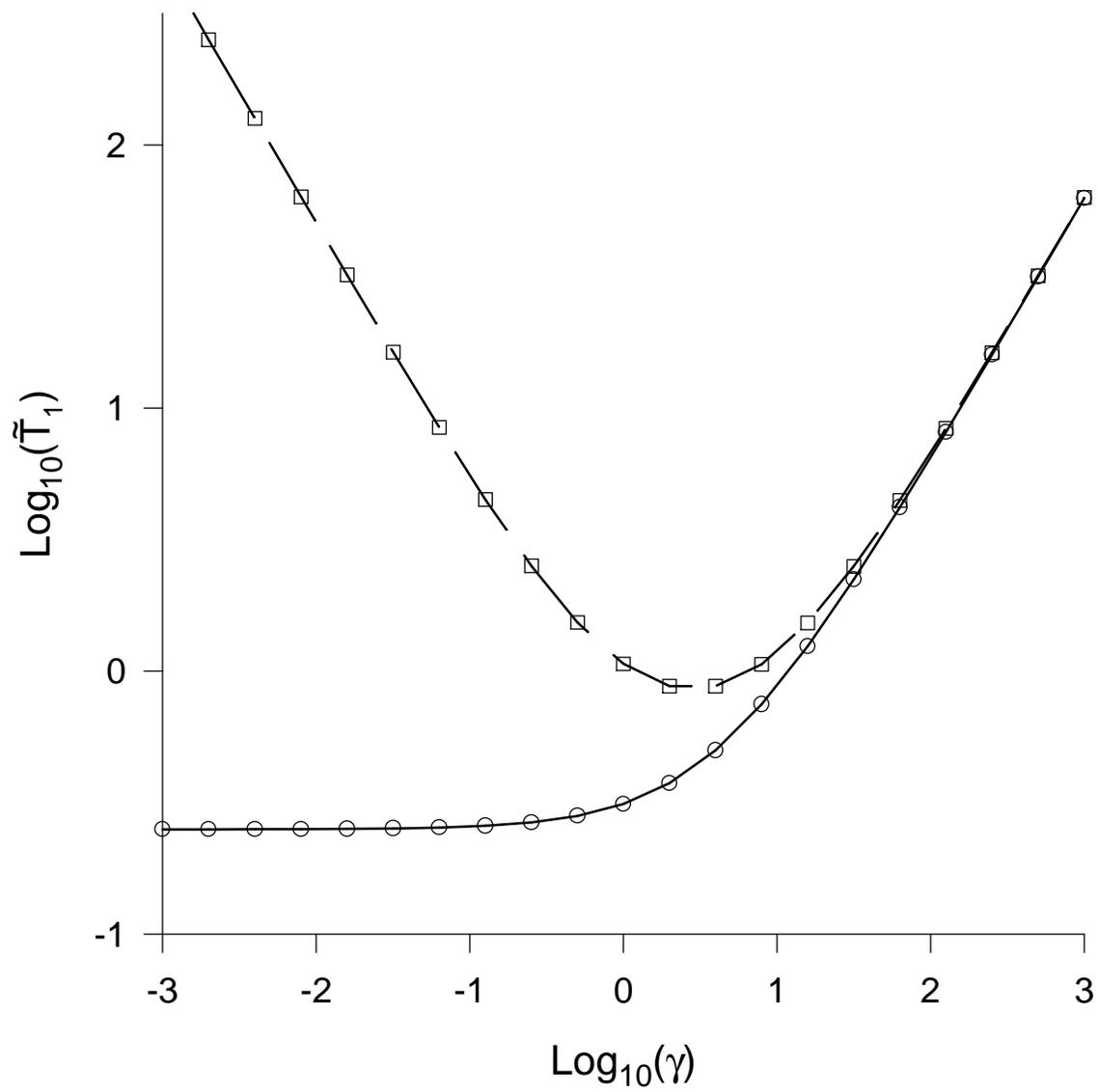
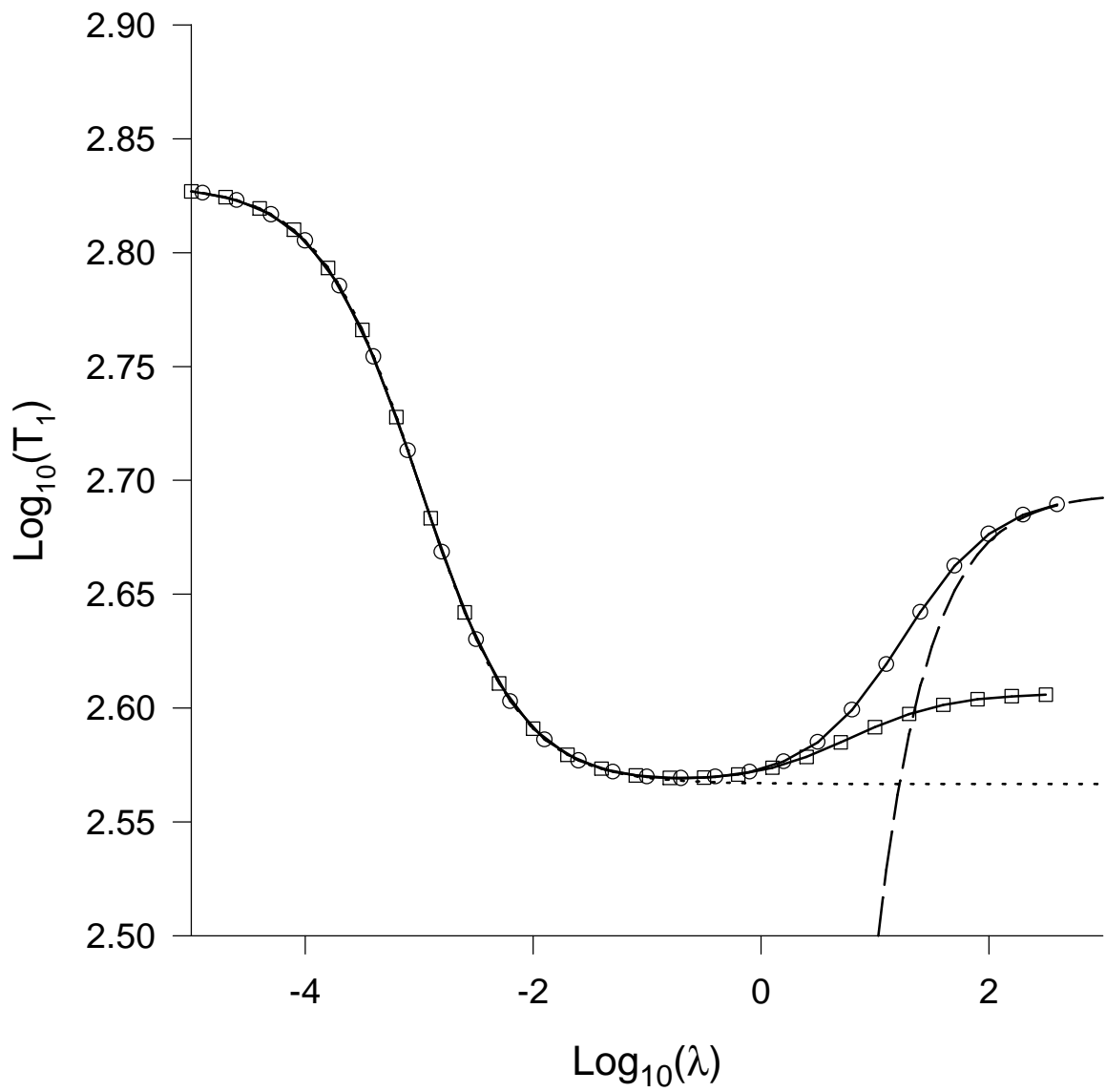
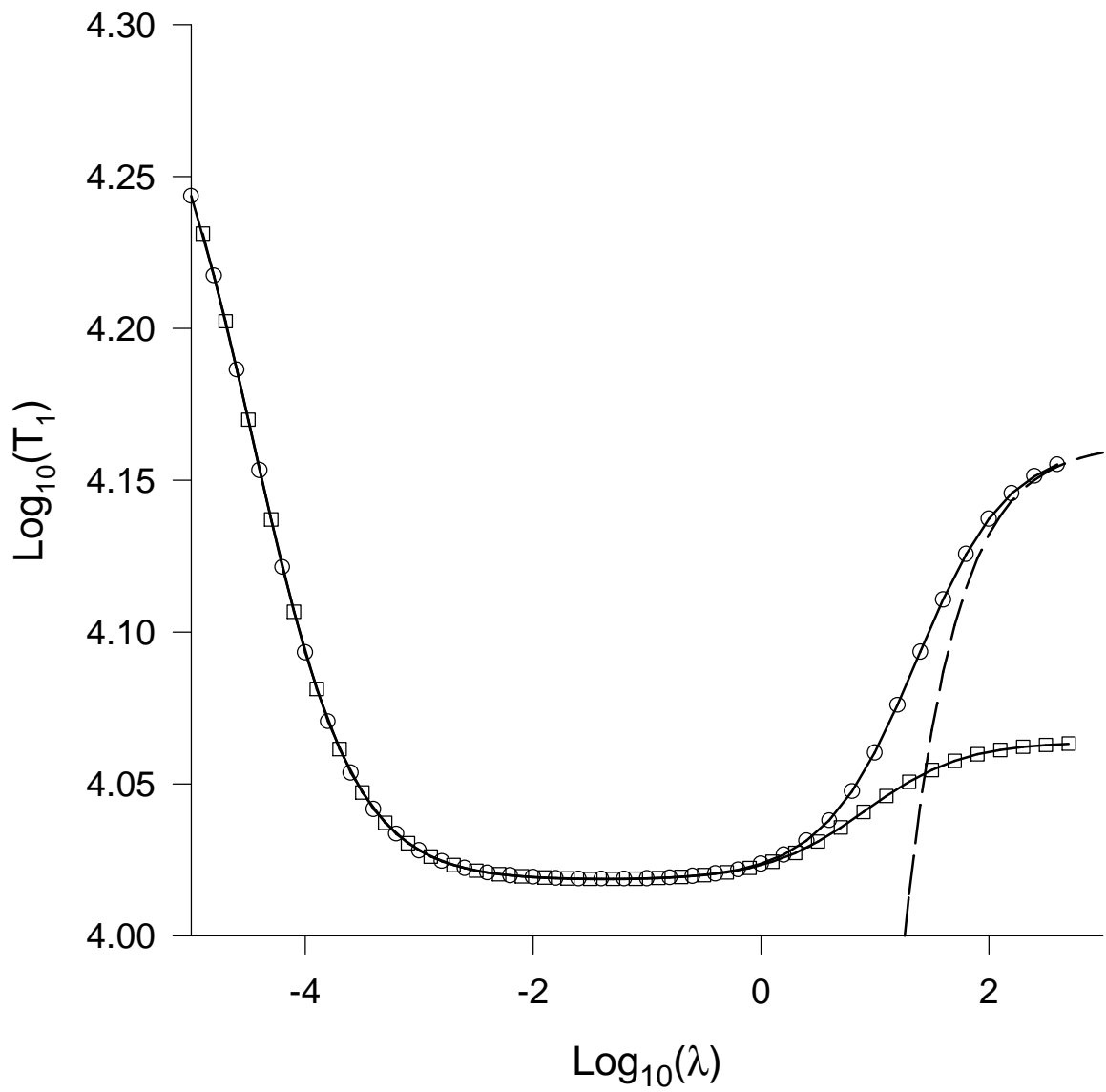
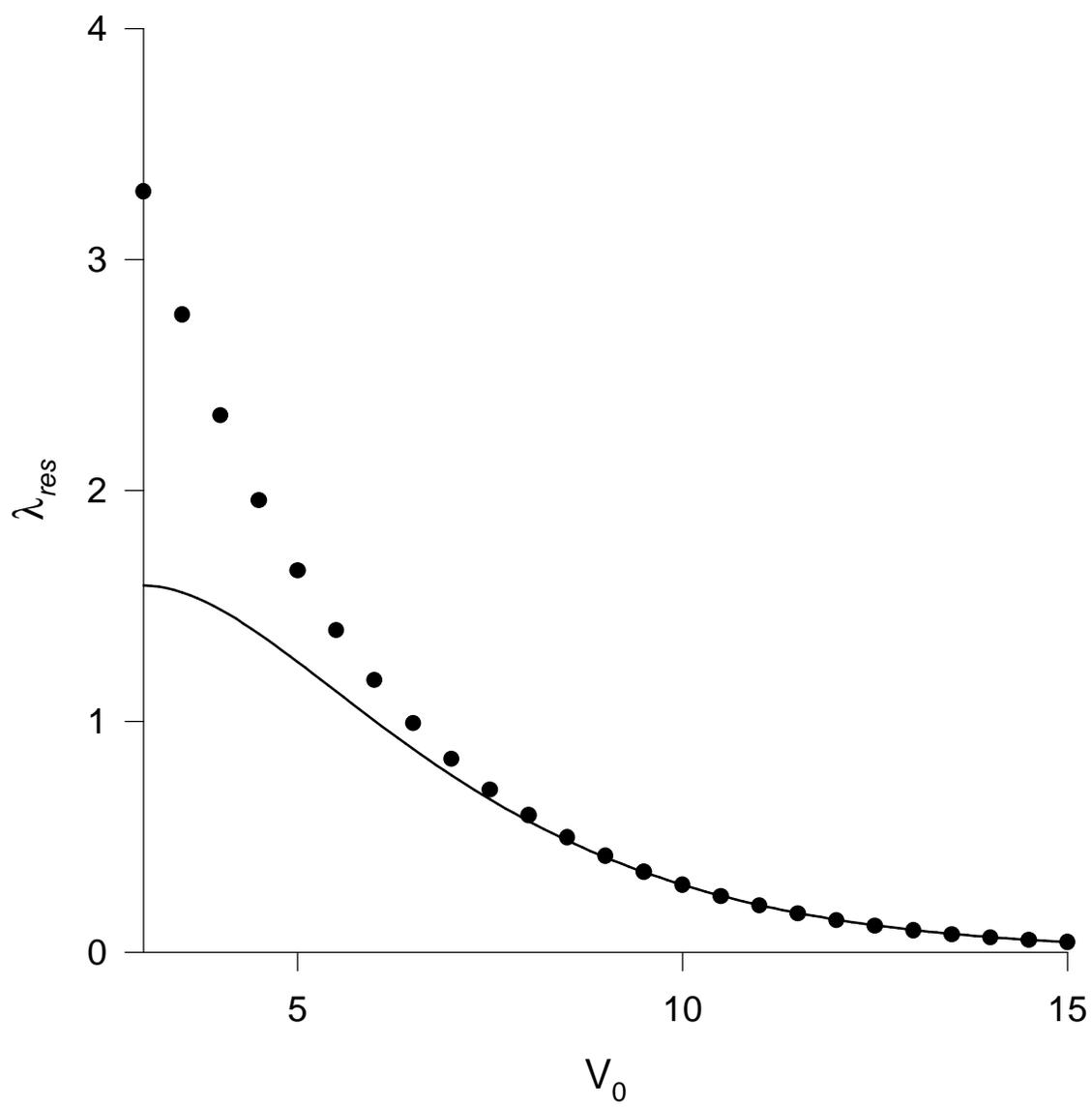
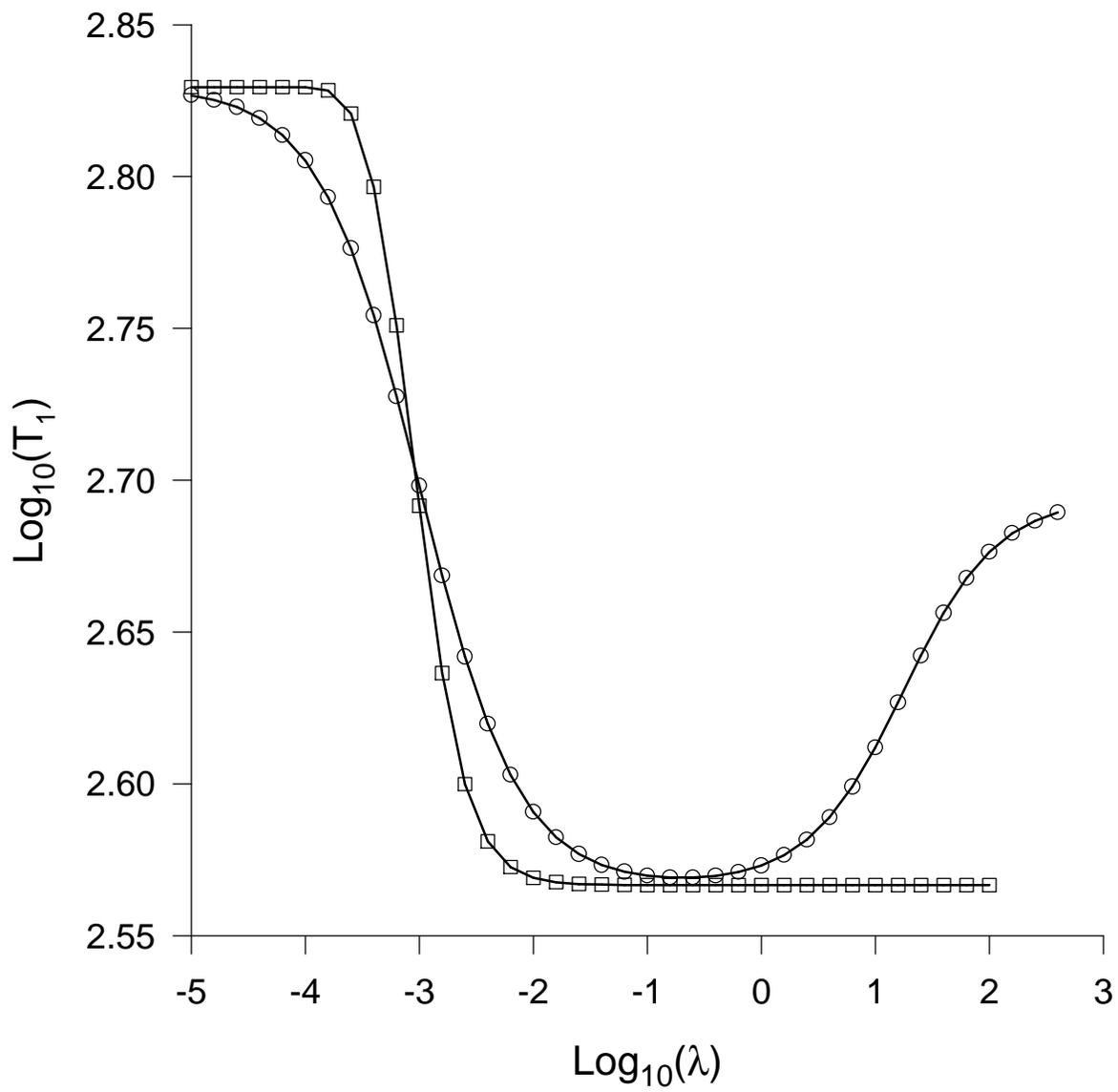
\end{document}